\documentclass[aps,pre,preprint,showkeys]{revtex4-1}
\usepackage{lipsum}
\usepackage[T1]{fontenc}
\usepackage[utf8]{inputenc}
\usepackage{amssymb,bm,setspace}
\usepackage{amsmath,amsfonts,relsize}
\usepackage{graphicx,hyperref,xcolor}
\usepackage{lineno,comment,amssymb}

\begin{document}
\title{Fisher information analysis of a Langevin System with Thermal and Athermal Noise} 
\author{E. S. Nascimento}
 \email{esnascimento@alumni.usp.br}
 \affiliation{Instituto Federal de Educação, Ciência e Tecnologia do Ceará, Campus Cedro, Alameda José Quintino, s/n, 63400-000, Cedro, CE, Brazil}
\author{W. A. M. Morgado }
 \email{welles@puc-rio.br}
 \affiliation{Departamento de Física, PUC-Rio, Rua Marquês de São Vicente 225, 22451-900, 
 Rio de Janeiro, RJ, Brazil \\ National Institute of Science and Technology - Complex Systems}

%
%
%
%
%
\begin{abstract}
Fisher Information (FI) has proven to be a useful tool for estimating the parameters of a system when only statistical data are available. We apply this method to a system of Langevin equations analogous to the Brownian Gyrator, in which the coupling parameter $b$ and the athermal noise strength $D$ are treated as unknown quantities. Using the FI, we discuss the estimation of the coupling parameter between the degrees of freedom and the quantity $D$ associated with the external noise effects. We find that the diagonal components of the FI exhibit nontrivial behavior, whereas the off-diagonal components display a monotonic dependence on the parameters of interest. This distinction may contribute to an accurate estimation of $b$ and $D$. 
\end{abstract}
%
%
\maketitle
%
%
%
\section{Introduction} \label{Intro}

Information is a quantity whose role in understanding the behavior of thermal systems has become increasingly important~\cite{Shannon1948,Parrondo2015}. Alternatively, information can be acquired through the estimation of parameters that cannot be measured directly. Since information acquisition is ultimately implemented through physical processes, its precision can be constrained by the thermodynamic costs associated with measurement and information  encoding~\cite{Granger2011,Micadei2013,Linpeng2022,Jacobs2012,Sagawa2010,Sagawa2013,Mancino2018,Pietzonka2024}. For example, the fairness of a die may be estimated by rolling it a large number of times, and testing the statistical significance of the null hypothesis of it being equiprobable. Fisher Information (FI)~\cite{Fisher1925} was introduced as a suitable quantity for measuring the amount of information that can be extracted about hidden parameters, being related to the inverse of the variance of their estimators. A comprehensive discussion of these techniques can be found in Ref.~\cite{JARice_livro,Kay_livro}.

In fact, we may assume that only some of the parameters that characterize a system are known, such as the mass and friction coefficient, while others remain unknown. Let us further assume that information about the system can be obtained through measurements, for example, by determining the moments of the position and velocity of a Brownian particle. However, the system may be influenced by distinct effective fluctuating forces acting along different directions. Such an asymmetry is known to arise in systems with dissipation and asymmetric energy injection, as occurs, for instance, in granular shaking experiments~\cite{Feitosa2002}. This asymmetry is characterized by an additional parameter $D$, which may be associated with the difference between effective temperatures. Another important hidden parameter is the dynamical coupling parameter $b$, which couples the dynamics along the two directions. From the probability distribution $P(x_1,x_2;b,D)$, obtained experimentally or analytically in the present case, we may construct the FI matrix~\cite{Melo2025Geometric,Melo2025SFI} that provides estimates for the variances and covariances of the parameters $(b,D)$.

The inverse FI matrix provides a lower bound on the covariance matrix of unbiased estimators. Through the Cram\'er--Rao bound~\cite{Cramer_livro,Rao1945}, one can determine the minimum possible variance of an estimator, that is, the smallest achievable estimation error. Therefore, when studying complex systems governed by several parameters, some of which are unknown, the FI becomes a useful tool for parameter estimation. Extracting implicit information generally involves constructing an estimator, proposing a value for the parameter, testing it against the available data, and finally determining a confidence interval. Consequently, the FI provides an estimate of the inverse variance associated with the deviation between the estimated and true values of a hidden parameter. The smaller the variance, the larger the FI.

The system considered in the present work is analogous to the Brownian Gyrator~\cite{FilligerReimann2007,ArgunVolpe2017,ItoDechant2020,Nascimento2021,Nascimento2022}, in which a Brownian particle is confined by a quadratic potential, and driven by distinct noise sources associated with reservoirs at different effective temperatures. As discussed above, the parameter $D$ characterizes the asymmetry arising from the distinct noise intensities in these directions, while the parameter $b$ characterizes the dynamical coupling between them. By observing the Brownian-particle dynamics and analyzing the corresponding FI matrix, we obtain information about the unknown parameters $(b,D)$. Applying the FI formalism enables us to investigate how these parameters influence both the dynamics and the stationary states of the system.

This paper is organized as follows. In Section II, we present the stochastic model of interest. In Section III, we discuss the covariance matrix and the probability distribution function. The analysis of the Fisher Information matrix is developed in Section IV. The discussion is presented in Section V. Finally, our conclusions are presented in Section VI.

\section{Stochastic differential equations} \label{2DLangevin}
Consider a Brownian particle moving in a two-dimensional space under the influence of deterministic and stochastic forces. The Cartesian coordinates of the particle position are given by $x_1$ and $x_2$. The particle is influenced by the confining quadratic potential
\begin{equation} \label{eq:potential_energy}
 U \left( x\right) = \frac{1}{2} x^\mathsf{T} K\, x, \quad K = 
 \begin{pmatrix}
  a & b \\
  b & a
 \end{pmatrix},
\end{equation}
where $a$ and $b$ are real quantities,  $a>0$, $a^2 > b^2$, and $x =  \left(x_1,x_2 \right)^\mathsf{T}$, with $\left(\cdot\right)^\mathsf{T}$ being the transpose of a matrix or vector.

\begin{figure}[h]
    \centering
    \includegraphics[scale=0.7]{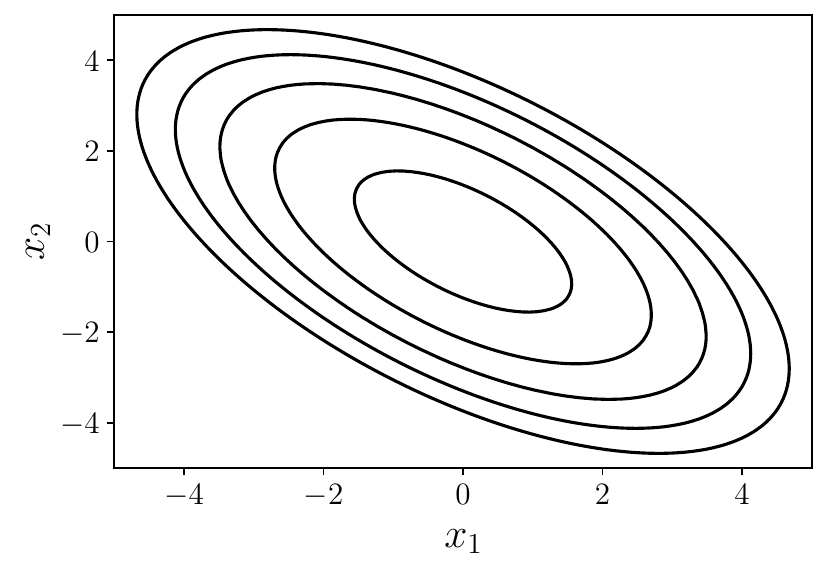}
    \caption{Contour plot of the potential energy $U \left( x_1, x_2 \right)$ on the $x_1 - x_2$ plane for $b = 0.6$.}
    \label{fig:potential_energy}
\end{figure}

We show in Fig.~\ref{fig:potential_energy} a contour plot of $U \left( x_1, x_2 \right)$ in the plane spanned by the coordinates $x_1$ and $x_2$. For $b \neq 0$, the principal axes of the potential $U$ are rotated about the origin of the Cartesian coordinate system. In addition to the linear restoring force associated with \eqref{eq:potential_energy}, there are damping and fluctuating forces related to noisy environments.

%
%

The dynamics of the model is described by the coupled overdamped Langevin equations
\begin{equation} \label{LanSys1}
\begin{split}
\gamma \dot{x}_1 \left( t \right) &=  - \frac{\partial U}{\partial x_1} + \eta_1 \left( t \right), \\
\gamma \dot{x}_2 \left( t \right) &=  - \frac{\partial U}{\partial x_2} + \eta_2 \left(t\right) + \xi_2 \left(t\right),  
 \end{split}
\end{equation}
with $x_i \left( 0 \right) = 0$, where $\gamma$ is the damping coefficient, $\eta_i \left(t\right)$ and $\xi_2 \left(t\right)$ are Gaussian noises, which are assumed to be independent. For simplicity, we consider a system of units where
\begin{equation} \label{eq:system_units}
a = \gamma = 1.
\end{equation}
Observe that the choice \eqref{eq:system_units} leads to
\begin{equation}\label{eq:stability_condition}
    -1<b<1
\end{equation}
for the stability condition of the potential energy \eqref{eq:potential_energy}.

The white Gaussian noises $\eta_1$ and $\eta_2$ represent the fluctuating effects of a thermal bath with temperature $T$, and $\xi_2 \left(t\right)$ is an external noise with strength parameter $D$, associated with a non-thermal reservoir. The cumulants of the Gaussian noises are \cite{van_Kampen-book,Gardiner}
\begin{equation} \label{noise_cumulants1}
 \langle \eta_{i}(t) \rangle_c = 0,\,  \langle \xi_{2}(t) \rangle_c= 0,
\end{equation}
\begin{equation} \label{noise_cumulants2}
 \langle \eta_{i}(t_1)\, \eta_{j}(t_2) \rangle_c = 2\, T \delta_{ij} \delta\left( t_1 - t_2 \right),\, \langle \xi_{2}(t_1)\, \xi_{2}(t_2) \rangle_c = 2\, D  \delta\left( t_1 - t_2 \right),
\end{equation}
\begin{equation} \label{noise_cumulants3}
  \langle \eta_{i}(t_1) \, \xi_{2}(t_2) \rangle_c = 0,
\end{equation}
where $\delta_{ij}$ is the Kronecker delta function, $\delta\left( t \right)$ is the Dirac-delta distribution. The idea here is to use the external noise to simulate a nonequilibrium situation with distinct effective temperatures, as discussed in \cite{MestresRoldan2014,DinisRica2016,ArgunVolpe2017}. Notice that the Boltzmann constant is assumed to be one. Also, without loss of generality, we may consider the following
\begin{equation}\label{eq:unit_temp}
 T \equiv 1.
\end{equation}
In fact, it is possible to make changes to the variables and parameters that lead to equations where all the quantities are dimensionless, which is more appropriate to work with mathematically.

The overdamped Langevin dynamics \eqref{LanSys1} can be written in matrix form
\begin{equation} \label{MatrixFormLangevin}
 \dot{x}\left( t \right) = -K\, x\left( t \right) + \zeta\left( t \right), 
\end{equation}
where
\begin{equation}
 x \left( t \right) = \begin{pmatrix}
 x_1 \left( t \right) \\
 x_2 \left( t \right)
 \end{pmatrix}, \quad 
 \zeta \left( t \right) = \begin{pmatrix}
 \eta_1 \left(t \right) \\
 \eta_2 \left(t \right) + \xi_2 \left(t \right)
 \end{pmatrix}, \quad 
  K = \begin{pmatrix}
  1 & b \\
  b & 1
 \end{pmatrix}.
\end{equation}
The eigenvalues of $K$ are $k_{\pm} = 1 \pm b$; 
the respective right eigenvectors are $ r_{+} = \left( 1, 1 \right)^\mathsf{T}$ and $r_{-} = \left( 1, -1 \right)^\mathsf{T}$.
Since $ -1< b< 1$, the eigenvalues $k_{\pm}$ are positive, which leads to stable solutions in the long-time limit. Then, the formal solution of \eqref{MatrixFormLangevin} reads
\begin{equation} \label{Formal_Solution_X}
 x\left( t\right) = \int_{0}^{t}  \exp \left[ - \left( t - s \right) K \right] \zeta\left( s \right) ds,
\end{equation}
with
\begin{equation}
\exp \left( - s K \right) = e^{-s}
\begin{pmatrix}
 \cosh \left( bs \right) & -\sinh \left( bs \right) \\
 -\sinh \left( bs \right) & \cosh \left( bs \right)
\end{pmatrix}.
\end{equation}

\section{Covariance matrix and probability distribution}

The linear model system \eqref{MatrixFormLangevin} is driven by independent Gaussian noises $\eta_1\left(t \right)$, $\eta_2\left(t \right)$, and $\xi_2\left(t \right)$. As a result, it is possible to explicitly obtain the elements $C_{ij} = \left< x_i x_j \right>_c$ of 
the covariance matrix $C$ associated with $x$ \cite{van_Kampen-book,Gardiner}. Then,
\begin{equation} \label{eq:C11}
 \begin{split}
  C_{11} &= \frac{1}{1 - b^2} \left\{ 1 - e^{-2t} \left[ b \sinh\left( 2bt \right) + \cosh\left( 2bt \right) \right]  \right\}
  + \\
 & \quad  \frac{D}{2\left( 1 - b^2 \right)} \left\{ b^2\left( 1 - e^{-2t} \right) + e^{-2t} \left[ b\sinh\left( 2bt \right) + 
 \cosh\left( 2bt \right) -1 \right]\right\},
 \end{split}
\end{equation}
\begin{equation} \label{eq:C12}
  C_{12} = C_{21} = \frac{2 + D}{2\left( 1 - b^2\right)} \left\{ e^{-2t} \left[ b\cosh\left( 2bt \right) + \sinh\left( 2bt \right)  \right] - b\right\},
\end{equation}
\begin{equation} \label{eq:C22}
 \begin{split}
  C_{22} &= \frac{1}{1 - b^2} \left\{ 1 - e^{-2t} \left[ b \sinh\left( 2bt \right) + \cosh\left( 2bt \right) \right]  \right\}
  + \\
 & \quad  \frac{D}{2\left( 1 - b^2 \right)} \left\{ 2 - b^2 + e^{-2t} \left[ b\sinh\left( 2bt \right) + 
 \cosh\left( 2bt \right) -1 \right]\right\},
 \end{split}
\end{equation}
The long-time limit of $C_{ij}$ reads
\begin{equation}\label{eq:sigma11}
 \begin{split}
  \sigma_{11} = \lim_{t \to \infty } C_{11} = \frac{2 + b^2 D}{2\left( 1-b^2 \right)},
 \end{split}
\end{equation}
\begin{equation}\label{eq:sigma12}
 \begin{split}
  \sigma_{12} =  \sigma_{21} = \lim_{t \to \infty } C_{12} = -\frac{b\left( 2 + D \right)}{2\left( 1-b^2 \right)},
 \end{split}
\end{equation}
\begin{equation}\label{eq:sigma22}
 \begin{split}
  \sigma_{22} = \lim_{t \to \infty } C_{22} = \frac{2 + \left(2- b^2 \right) D}{2\left( 1-b^2 \right)}.
 \end{split}
\end{equation}

Since $x$ is a Gaussian stochastic variable with covariances \eqref{eq:C11}-\eqref{eq:C22}, one may write explicitly the probability distribution $P\left( x,t;b,D  \right)$ of position variable $x$ at time $t$, for a given value of coupling parameter $b$ and external noise strength $D$. Therefore,
\begin{equation} \label{eq:prob_density1}
 \begin{split}
  P\left( x,t;b,D \right) = \frac{1}{\sqrt{4\pi^2\,\text{det} \left(C\right)}} \exp\left( - \frac{1}{2} x^\intercal C^{-1} x \right),
 \end{split}
\end{equation}
where $\text{det} \left(C\right)$ is the determinant of $C$. In the long-time limit ($t \to \infty$),  the steady-state probability distribution $P_{st} \left(x; b,D \right)$ is given by
\begin{equation} \label{eq:stat_dist}
 P_{st} \left(x; b,D \right) =  \frac{1}{\sqrt{4\pi^2\,\text{det} \left(\sigma\right)}} \exp\left( - \frac{1}{2} x^\intercal \sigma^{-1} x \right),
\end{equation}
where $\sigma$ is the stationary covariance matrix with elements \eqref{eq:sigma11}-\eqref{eq:sigma22}. 

\begin{figure}[h]
 \centering
 \includegraphics[scale=0.8]{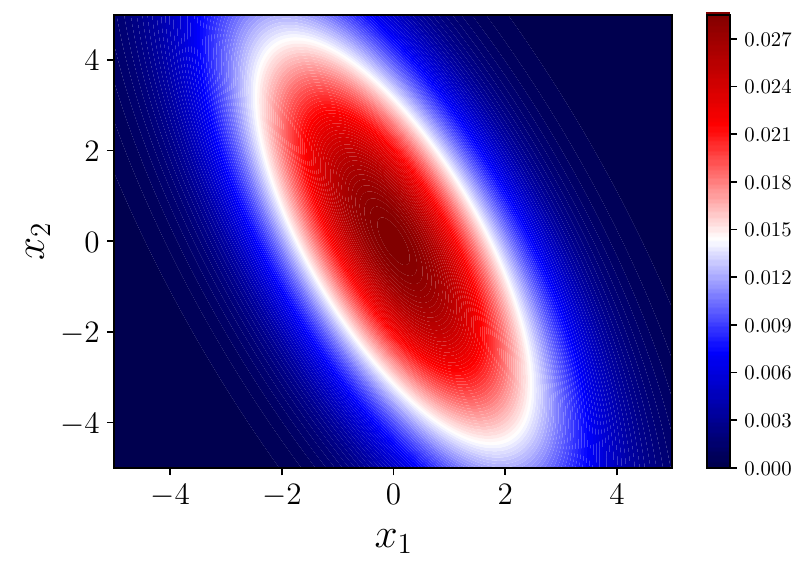}
 \caption{Contour plot of the stationary probability distribution of position variables for $b=0.6$ and $D = 10$.}
 \label{fig:stat_prob_dist_plot}
\end{figure}

Figure \ref{fig:stat_prob_dist_plot} shows the contour plot of $P_{st} \left(x; b,D \right)$ in the plane spanned by the coordinates $x_1$ and $x_2$ for $b = 0.6$ and $D = 10$. The presence of distinct noises along the directions $x_1$ and $x_2$ leads to an asymmetry of the distribution $P\left(x_1,x_2;b,D\right)$ compared to the potential energy $U\left(x_1,x_2\right)$ shown in Fig.~\ref{fig:potential_energy}. This behavior is reminiscent of the Brownian Gyrator \cite{FilligerReimann2007}, a stochastic system that has received attention due to its interesting non-equilibrium properties \cite{ArgunVolpe2017,Nascimento2021,BaeJeong2021,Squarcini_2022}. Ciliberto \emph{et al.} \cite{CilibertoImparatoTanaseNaert2013} investigated experimentally an analogous system  consisting of a  capacitor-resistor circuit with random voltages and different temperatures associated with thermal noise. 

\section{Fisher Information about $b$ and $D$}

The function $P\left( x,t;b,D\right)$ gives the probability density of finding the system state $x = \left(x_1,x_2 \right)$ at time $t$ for a given value $b$ and $D$. However, in many real-world situations, the parameters $b$ and $D$ are generally not known \textit{a priori} and must be inferred from measurements. Then, it is necessary to determine the amount of information about these parameters from the observed dynamical behavior of the system. In fact, it is possible to do that by means of the Fisher Information (FI) matrix \cite{Miller_livro, Kay_livro},
\begin{equation}\label{eq:FIM}
 I = \begin{pmatrix}
     I_{11} & I_{12} \\
     I_{21} & I_{22}
 \end{pmatrix} ,  
\end{equation}
with
\begin{equation}\label{eq:app_FI}
I_{ij} =
\left\langle
\left(\frac{\partial}{\partial\theta_i}\ln P\right)
\left(\frac{\partial}{\partial\theta_j}\ln P\right)
\right\rangle,
\end{equation}
where $\theta_1$ and $\theta_2$ are the model parameters of interest, and the average is taken with respect to the probability density $P$. Notice that the FI matrix is symmetric, so that $I_{12}=I_{21}$.

The Fisher Information quantifies the sensitivity of the probability distribution to variations of the parameters $(b,D)$. The connection between Fisher Information and parameter estimation is established by the Cram\'er--Rao bound~\cite{Rao1945,Cramer_livro}. For an unbiased estimator $\hat{\boldsymbol{\theta}}$ of the parameter vector $\boldsymbol{\theta}$, and under the appropriate regularity conditions, the covariance matrix satisfies $\operatorname{Cov}(\hat{\boldsymbol{\theta}})
\succeq
I^{-1}(\boldsymbol{\theta})$. Thus, a larger FI generally corresponds to the possibility of estimating the parameters with greater precision. 

In the present work, rather than constructing estimators from experimental data, we calculate the Fisher Information directly from the theoretically determined probability density $P(x,t;b,D)$. This allows us to quantify how much information about the coupling $b$ and the external noise strength $D$ is encoded in the statistical state of the system.

For the zero-mean Gaussian distribution \eqref{eq:prob_density1}, it is possible to show that~\cite{Kay_livro,Miller_livro,MalagoPistone2015}
\begin{equation}\label{eq:eq1_FI}
I_{ij} = \frac{1}{2}\operatorname{tr}
\left(
C^{-1}
\frac{\partial C}{\partial\theta_i}
C^{-1}
\frac{\partial C}{\partial\theta_j}
\right),
\end{equation}
where $\operatorname{tr}(\cdot)$ denotes the trace of a matrix,
\begin{equation}
     \theta_1 = b, \quad  \theta_2 = D,
\end{equation}
and $\partial C/\partial\theta_i$ is a matrix with elements
$\partial C_{mn}/\partial\theta_i$.
The diagonal elements $I_{11}$ and $I_{22}$ quantify the sensitivity of the probability distribution to variations in $b$ and $D$, respectively, while the off-diagonal element $I_{12}=I_{21}$ characterizes the statistical coupling between the corresponding score functions.

It is worth mentioning that, for the Brownian Gyrator type model considered here, we are particularly interested in the FI associated with the parameters related to the nonequilibrium behavior of the system. Also, the elements $I_{ij}$ are time-dependent quantities, since the covariance matrix $C$ evolves in time. Consequently, the amount of information about $b$ and $D$ encoded in the instantaneous probability distribution may change during the evolution of the system.

\subsection{Time-dependent properties of Fisher Information}

\begin{figure}[h] 
 \centering 
 \includegraphics[scale=0.45]{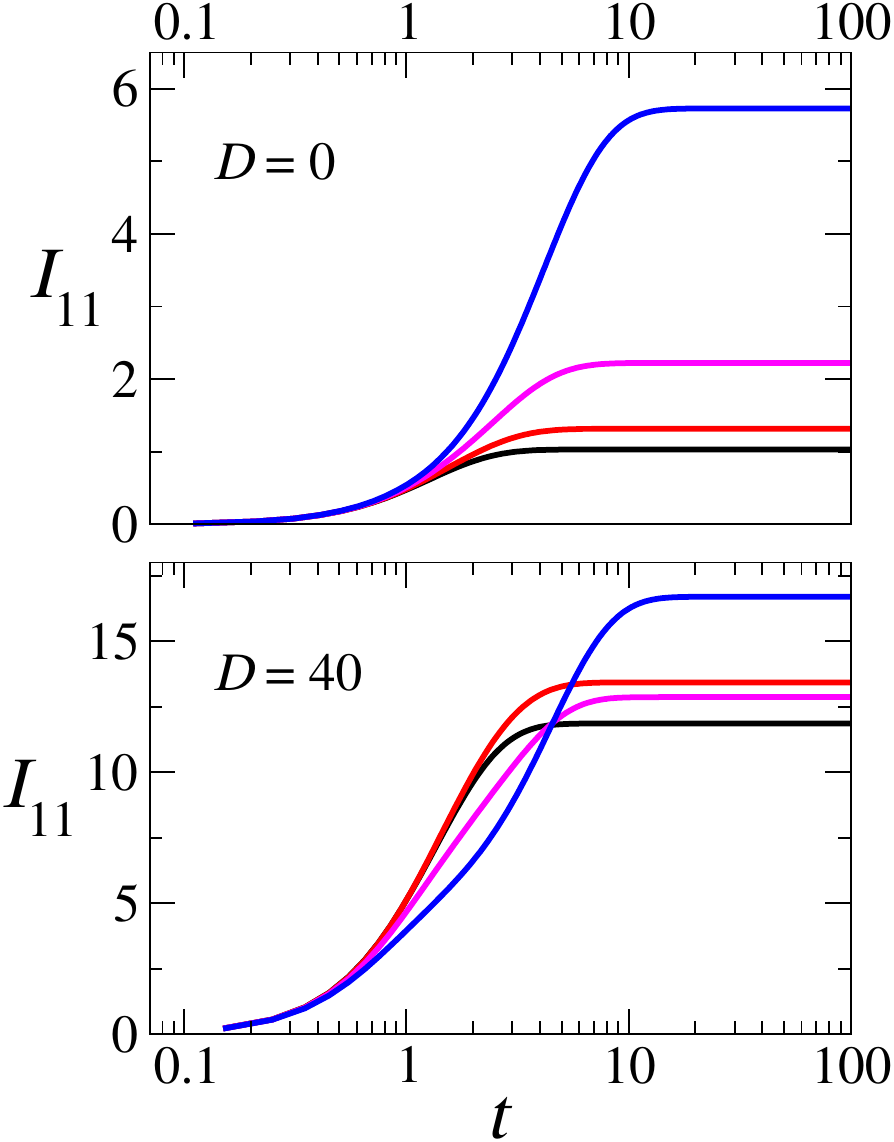} 
 \caption{Element of FI matrix $I_{11}$ \emph{versus} dimensionless $t$ for external noise strength $D = 0$ and $D = 40$. Values of coupling parameter $b$: 0.1 (black), 0.3 (red), 0.5 (magenta) and 0.7 (blue). } 
 \label{fig:I11_vs_times} 
\end{figure}

Figure \ref{fig:I11_vs_times} shows the FI matrix element $I_{11}$ as a function of the dimensionless time $t$ for different values of the coupling parameter $b$ and external noise strength $D$. The quantity $I_{11}$ is an even function of $b$, i.e., $I_{11}(b,D,t)=I_{11}(-b,D,t)$. Therefore, we restrict the discussion to non-negative values of $b$.

The element $I_{11}$ is typically zero near $t=0$, undergoes a transient increase, and eventually approaches a stationary value. A clear dependence on both $b$ and $D$ can be observed. For $D=0$ and a given value of $b$, the FI $I_{11}$ increases with time, approaching a constant value as $t\to\infty$. In addition, its long-time limit increases with $b$, indicating that the probability distribution becomes increasingly sensitive to variations of the coupling parameter. As $|b|$ increases, the coupling has a stronger effect on the covariance structure of the coordinates $x_1$ and $x_2$, which changes the amount of information about $b$ encoded in their joint probability distribution.

For small fixed values of $D$, $I_{11}$ increases with $b$, similarly to the case $D=0$. However, as shown in Figure \ref{fig:I11_vs_times}, for sufficiently large fixed values of the external noise strength, $I_{11}$ may exhibit a non-monotonic dependence on $b$. This behavior is also found in the long-time limit. Thus, in the presence of sufficiently strong external noise, increasing the coupling does not necessarily increase the information about $b$ contained in the probability distribution.

\begin{figure}[h]
 \centering
 \includegraphics[scale=0.45]{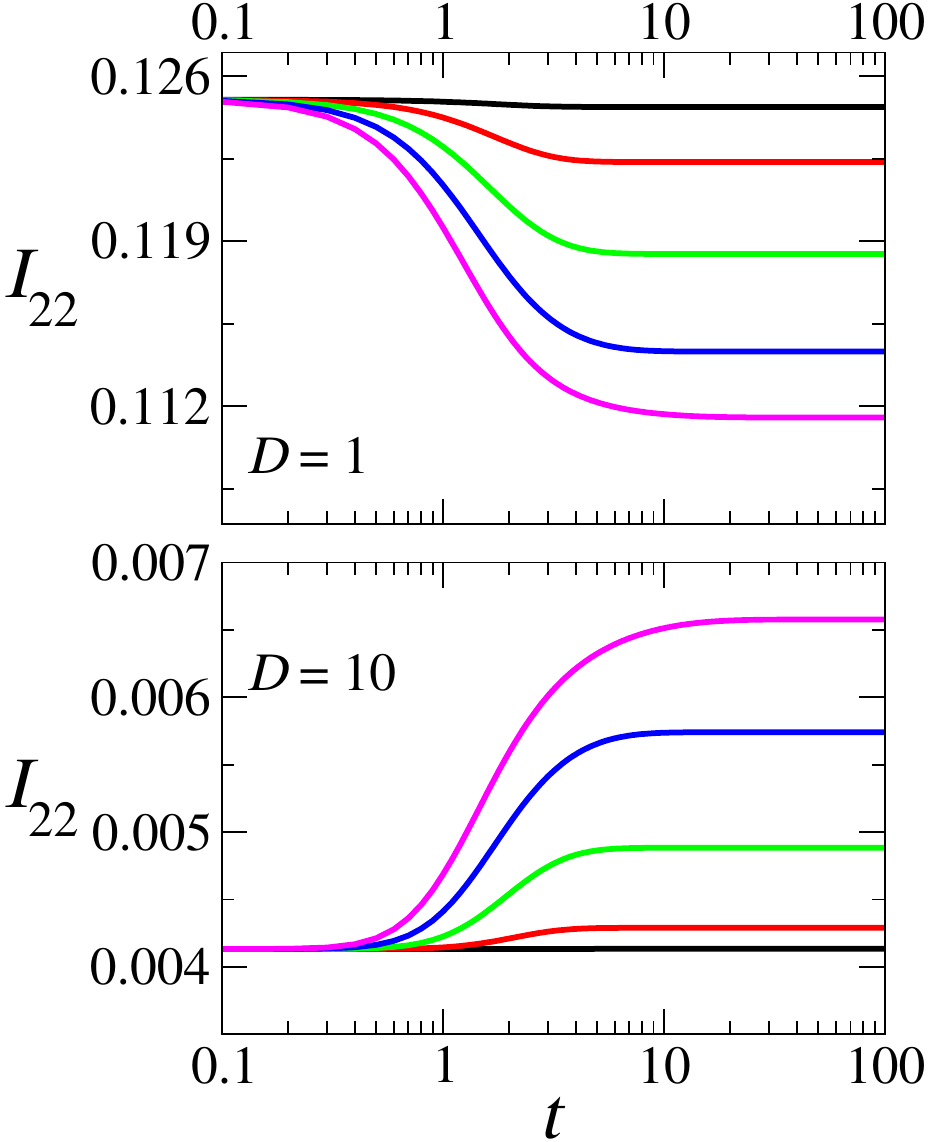}
 \caption{Element of FI matrix $I_{22}$ \emph{versus} dimensionless $t$ for $D = 1$ and $D = 10$. Values of $b$:  0.1 (black),  0.3 (red), 0.5 (green), 0.7 (blue), and 0.9 (magenta). }
 \label{fig:I22_vs_times}
\end{figure}

In Figure \ref{fig:I22_vs_times}, we show the FI matrix element $I_{22}$ as a function of the dimensionless time $t$, for $D=1$ and $D=10$ and different values of $b$. In general, $I_{22}$ decreases as $D$ increases, as can also be seen from the different vertical scales in Figure \ref{fig:I22_vs_times}. Since $I_{22}=I_{DD}$ measures the sensitivity of the probability distribution to variations of $D$, this behavior indicates that the distribution becomes relatively less sensitive to further changes in the external noise strength as $D$ increases. The coupling parameter $b$ also affects $I_{22}$, showing that the amount of information about the external noise strength depends on the coupling between the two coordinates.

\subsection{FI for the stationary distribution}

Consider the stationary probability distribution shown in \eqref{eq:stat_dist}. In this steady-state scenario, the time-independent FI matrix reads
\begin{equation}
   \mathcal{I} =  \begin{pmatrix}
        \mathcal{I}_{11} & \mathcal{I}_{12} \\
        \mathcal{I}_{12} & \mathcal{I}_{22} 
    \end{pmatrix},
\end{equation}
where
\begin{equation}\label{I_11_s}
 \mathcal{I}_{11} = \frac{\left( 2 + D\right)^2}{ \left( 1-b^2 \right)^2 } \left[  \frac{ 4 \left( 1 + D \right)\left( 1 + b^2 \right) + b^2 D^2 \left( 3 - b^2 \right) }{\left( 4 + 4D + b^2 D^2 \right)^2 } \right],
\end{equation}
\begin{equation}\label{I_12_s}
 \mathcal{I}_{12} = \frac{ b\left( 2 + D\right)  }{ 1 - b^2 } \left[ \frac{ 4 + b^2 D \left( 4 + D \right) }{  \left( 4 + 4D + b^2 D^2 \right)^2  }\right],
\end{equation}
\begin{equation}\label{I_22_s}
 \mathcal{I}_{22} = \frac{ 4 \left( 2 - b^2 \right) + b^2 D \left( 4 + b^2  D\right) }{ \left( 4 + 4D + b^2 D^2 \right)^2 }.
\end{equation}
Notice that $-1<b<1$ for the Langevin system considered here in order to achieve stable solutions for long times. The quantities $\mathcal{I}_{ij}$ are rational functions of the coupling parameter $b$ and the external noise strength $D$. Also, $\mathcal{I}_{11}$ and $\mathcal{I}_{22}$ are even functions of $b$ and $\mathcal{I}_{12}$ is an odd function of $b$. 

In the long-time limit, the determinant of FI matrix is
\begin{equation}\label{det_FIM}
\det \mathcal{I} = \mathcal{I}_{11} \mathcal{I}_{22} - \mathcal{I}_{12}^2 = \frac{(2+D)^2\,N(b,D)}{(1-b^2)^2\left(4+4D+b^2D^2\right)^3},
\end{equation}
where
\begin{equation}\label{N_bD}
N(b,D) = D^2b^4(2-b^2)
+
4Db^2(2-b^2)
+
4(2-b^4).
\end{equation}
For $D\geq 0$ and $|b|<1$, all the terms in $N(b,D)$ are non-negative. Moreover, since $2-b^2>0$ and $2-b^4>0$ in the stable region, one has $N(b,D)>0$. Therefore,
\begin{equation}
\det\mathcal{I}>0,
\quad D\geq0,\quad |b|<1.
\end{equation}

\begin{figure}[h]
 \centering
 \includegraphics[scale=0.8]{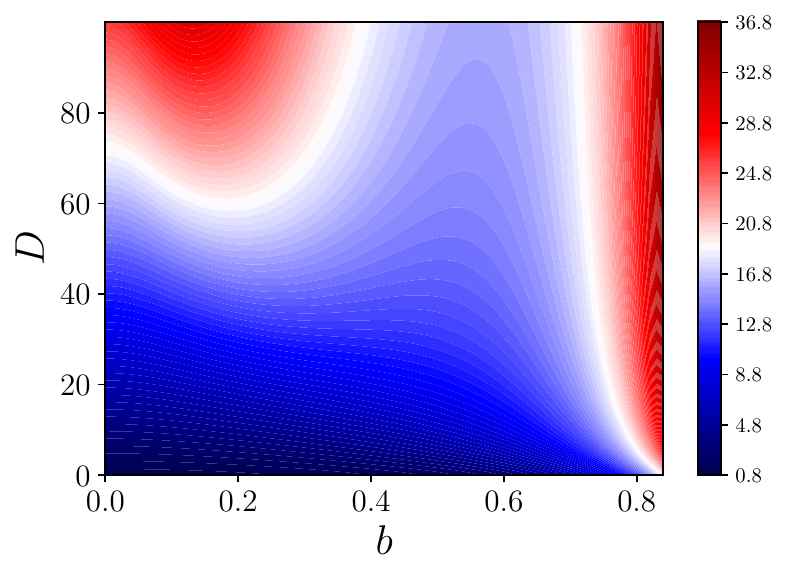}
 \caption{Contour plot of $\mathcal{I}_{11}$ on $b-D$ plane for stationary states.}
  \label{I_11_st}
\end{figure}

\begin{figure}[h]
 \centering
 \includegraphics[scale=0.8]{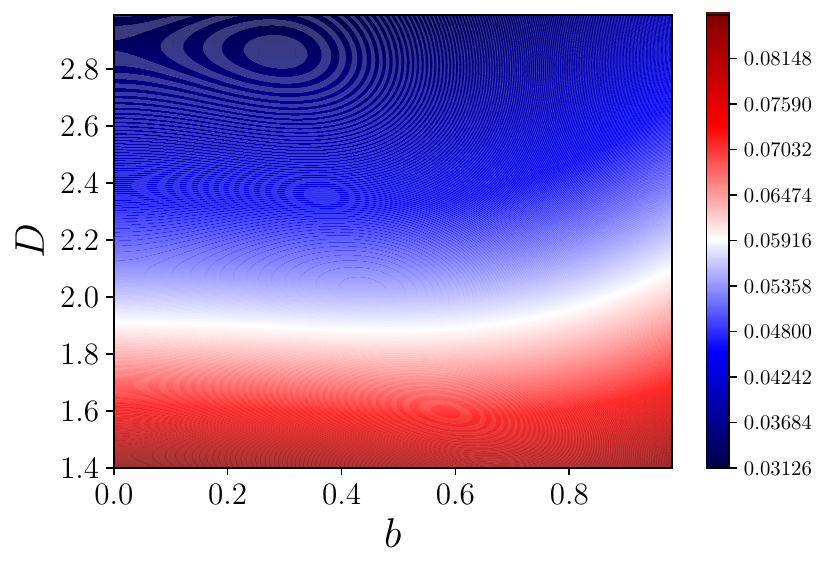}
 \caption{Contour plot of $\mathcal{I}_{22}$ on $b-D$ plane for stationary states.}
  \label{I_22_st}
\end{figure}

Since $\mathcal{I}_{11}>0$ and $\mathcal{I}_{22}>0$, the trace of $\mathcal{I}$ is positive,
\begin{equation}
    \textrm{tr}\, \mathcal{I} = \mathcal{I}_{11} + \mathcal{I}_{22} >0.
\end{equation}
Consequently, the FI matrix $\mathcal{I}$ is positive definite, as expected. We show in FIGs. \ref{I_11_st}-\ref{I_22_st} the contour plots of $\mathcal{I}_{ij}$ on the $b-D$ plane.

As previously discussed in this work, the element $\mathcal{I}_{11}$ exhibits a non-monotonic behavior as a function of the coupling parameter $b$ for sufficiently large fixed values of $D$. In fact, $\mathcal{I}_{11}$ increases with $D$ when $b$ is in a neighborhood of $0$ and $\pm 1$. In the vicinity of $b=0$, the Fisher Information element $\mathcal{I}_{11}$ can be expanded in powers of the coupling parameter $b$. Up to second order, one obtains
\begin{equation}
\mathcal{I}_{11} =
\frac{(D+2)^2}{4(D+1)}
+
\frac{(D+2)^2\left(D^2+12D+12\right)}
{16(D+1)^2}\,b^2
+
O\left(b^4\right).
\end{equation}
As expected from the symmetry $\mathcal{I}_{11}(b,D)=\mathcal{I}_{11}(-b,D)$, only even powers of $b$ appear in the expansion. Moreover, the coefficient of $b^2$ is positive for $D\geq0$, showing that, for any fixed value of $D$, $\mathcal{I}_{11}$ initially increases as $|b|$ moves away from zero. Then, in the vicinity of $b = 0$, $\mathcal{I}_{11}$ increases with $D$. On the other hand, from \eqref{I_11_s} it follows that $\mathcal{I}_{11}$ diverges as $b$ approaches $\pm 1$.

The element $\mathcal{I}_{22}$ generally decreases with $D$, indicating that the probability distribution becomes relatively less sensitive to further changes in the athermal noise strength. Similarly, in the vicinity of $b=0$, the Fisher Information element $\mathcal{I}_{22}$ can be expanded in powers of the coupling parameter $b$.  Up to second order, one obtains
\begin{equation}
\mathcal{I}_{22} = \frac{1}{2(D+1)^2} - 
\frac{1}{4(D+1)^3}\,b^2 + O\left(b^4\right). \label{eqexpCinD}
\end{equation}
As expected from the symmetry $\mathcal{I}_{22}(b,D)=\mathcal{I}_{22}(-b,D)$, only even powers of $b$ appear in the expansion. The coefficient of $b^2$ is negative for $D\geq0$, showing that, for any fixed value of $D$, $\mathcal{I}_{22}$ initially decreases as $|b|$ moves away from zero. Near $b=0$, the information about $D$ decreases as $|b|$ increases and also decreases as $D$ increases.

Notice that the covariance matrix $C$, with elements \eqref{eq:C11}-\eqref{eq:C22}, depends linearly on $D$. Indeed, one may write
\begin{equation}
C(t;b,D) = C^{(0)}(t;b)+ D\,C^{(D)}(t;b).
\end{equation}
For sufficiently large values of $D$,
\begin{equation}
C(t;b,D)
\sim
D\,C^{(D)}(t;b),
\quad
C^{-1}(t;b,D)
\sim
\frac{1}{D}
\left[C^{(D)}(t;b)\right]^{-1}, \quad \frac{\partial C}{\partial D}
\sim
C^{(D)}(t;b).
\end{equation}

Using the expression for the Fisher Information \eqref{eq:eq1_FI}, one therefore expects, in the large-$D$ limit,
\begin{equation}
I_{DD}
\propto
\frac{1}{D^{2}}.
\end{equation}
In this regime, the probability distribution becomes progressively less sensitive to changes in $D$, and consequently contains less Fisher Information about the external noise strength. Then, considering $b$ near $0$ or $\pm 1$, the element $\mathcal{I}_{22}$ decreases as $D$ increases, which is indicated in Fig.~\ref{I_22_st}.
\begin{figure}[h]
 \centering
 \includegraphics[scale=0.8]{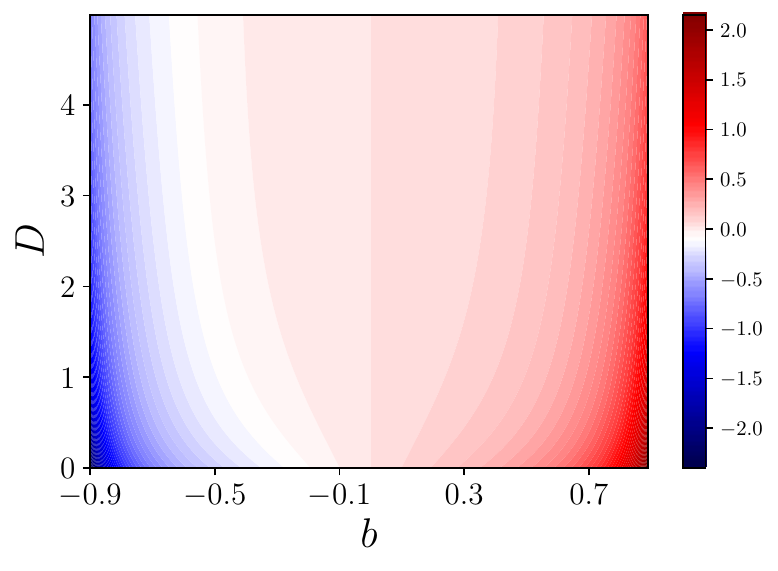}
 \caption{Contour plot of $\mathcal{I}_{12}$ on $b-D$ plane for stationary states.}
  \label{I_12_st}
\end{figure}

Again, in the vicinity of $b=0$, the off-diagonal Fisher Information element $\mathcal{I}_{12}$ can be expanded in powers of the coupling parameter $b$,
\begin{equation}
\mathcal{I}_{12} = \frac{D+2}{4(D+1)^2}\,b
+ O\left(b^3\right).
\end{equation}
Due to the antisymmetry $\mathcal{I}_{12}(b,D)=-\mathcal{I}_{12}(-b,D)$, only odd powers of $b$ appear in the expansion. In particular, $\mathcal{I}_{12}=0$ at $b=0$, while for small nonzero $b$ the off-diagonal Fisher Information increases linearly with $b$, its sign determined by the sign of $b$.

\section{Discussion}

\subsection{Information signatures of b and D}

The parameters $b$ and $D$ play very different physical roles in the model. The parameter $b$ determines the coupling between the two spatial directions and therefore controls the development of statistical correlations between the position variables $x_1$ and $x_2$. The quantity $D$ determines the strength of additional athermal noise acting directly in the $x_2$ direction. However, through the coupling, the effects of this additional noise can also propagate to $x_1$, modifying the covariance structure of the joint probability distribution.

Given the deterministic initial condition of the model, the initial probability distribution is independent of $b$ and $D$, and therefore carries no information about these parameters. As the system evolves, the dynamics progressively encodes information about $b$ and $D$ into the covariance structure of the probability distribution. This encoding process is observed for $I_{11}$ in Fig.~\ref{fig:I11_vs_times}. A similar behavior is observed for $I_{22}$ in Fig.~\ref{fig:I22_vs_times}. For $D=10$, the gradual increase towards the stationary value is clearly resolved, whereas for $D=1$ this process occurs on a considerably shorter time scale.

The behavior of FI $I_{11}$ and $I_{22}$ is clearly different. The FI for $b$ increases with $b$, even diverging at $|b|=1$. On the other hand, the FI of $D$ actually decreases as $D$ increases (with $I_{22}\sim D^{-2}$ as $D \gg 1$), showing that stronger athermal noise does not mean that its strength becomes easier to determine. In contrast, once \(D\) is large, the relative changes in \(D\) produce progressively less distinguishable changes in the distribution.

\subsection{FI divergence and boundary stability}

The divergence of $\mathcal{I}_{11}$ as $|b|\to1$ can be understood from the stability properties of the Langevin dynamics. The eigenvalues associated with Eq.~\eqref{MatrixFormLangevin} are $k_{\pm}=1\pm b$. Therefore, as $b$ approaches the stability boundary, one of the eigenvalues tends to zero, and an increasingly slow relaxation mode develops. At the same time, the stationary covariance matrix contains the factor $(1-b^2)^{-1}$ and becomes increasingly sensitive to variations of $b$. This enhanced sensitivity produces the divergence of the Fisher Information $\mathcal{I}_{11}$ as $|b|\to1$ from within the stable region.

\subsection{Off-diagonal information}

For the present model, $\mathcal{I}_{12}$ is an odd function of the coupling parameter $b$. In the vicinity of $b=0$,
\begin{equation}
\mathcal{I}_{12}
=
\frac{D+2}{4(D+1)^2}\,b
+
O\left(b^3\right).
\end{equation}
At $b=0$, the position variables $x_1$ and $x_2$ are uncoupled, and the Fisher Information matrix is diagonal. As soon as a finite coupling is introduced, an off-diagonal contribution develops linearly with $b$, indicating that the statistical signatures associated with variations of $b$ and $D$ become coupled. Furthermore, reversing the sign of $b$ reverses the sign of $\mathcal{I}_{12}$ while leaving the diagonal elements unchanged.

The effect of $\mathcal{I}_{12}$ on joint parameter estimation may be related to the behavior of $\det\mathcal{I}
=
\mathcal{I}_{11}\mathcal{I}_{22}
-
\mathcal{I}_{12}^{\,2}$.
A large value of $|\mathcal{I}_{12}|$ relative to $\sqrt{\mathcal{I}_{11}\mathcal{I}_{22}}$ indicates that the statistical responses to $b$ and $D$ are aligned, making their simultaneous estimation difficult. Conversely, when $\mathcal{I}_{12}$ is small, the two parameters affect the probability distribution in a nearly independent way. Therefore, $\mathcal{I}_{12}$ may play a role in determining how the coupling parameter $b$ and the external noise strength $D$ can be inferred from the stochastic state of the system.

\section{Conclusion} \label{Con}

In this work, we investigated the Fisher Information associated with two parameters governing the nonequilibrium behavior of a two-dimensional coupled Langevin system driven by thermal and athermal noises. The coupling parameter $b$ controls the interaction between the two degrees of freedom of interest, whereas $D$ determines the strength of the additional athermal noise present in the model. Since the stochastic model is linear and driven by Gaussian noise, the instantaneous probability distribution can be obtained analytically, allowing the Fisher Information matrix associated with $(b,D)$ to be calculated directly from the covariance matrix.

The time-dependent analysis shows how information about the model parameters is encoded in the probability distribution during the stochastic evolution. Starting from a parameter-independent deterministic initial condition, the Fisher Information develops during the transient regime and eventually approaches its stationary value. The two diagonal elements of the FI matrix display markedly different behaviors. The information $\mathcal{I}_{11}$ about the coupling $b$ generally increases as the stability boundary $|b|=1$ is approached, whereas the information $\mathcal{I}_{22}$ about the athermal noise strength decreases as $D$ becomes large. In particular, $\mathcal{I}_{22}\sim D^{-2}$ in the large-$D$ regime. Thus, increasing the strength of the athermal fluctuations does not necessarily improve our ability to determine their magnitude; instead, sufficiently strong noise makes further variations of $D$ less distinguishable from the statistical state of the system.

The increase of $\mathcal{I}_{11}$ near $|b|=1$ can be related to the stability properties of the Langevin dynamics. The relaxation timescales are inversely proportional to $k_{\pm}=1\pm b$, so that one relaxation mode becomes arbitrarily slow as the stability boundary is approached. Correspondingly, the stationary covariance becomes increasingly sensitive to variations of $b$, producing a divergence of the Fisher Information associated with $b$. The off-diagonal element $\mathcal{I}_{12}$ provides complementary information about the simultaneous inference of $b$ and $D$. It vanishes in the uncoupled case $b=0$ and develops linearly with $b$ for weak coupling. The  determinant of the FI matrix is positive throughout the stable region $|b|<1$, $D\geq0$, demonstrating that the two parameters remain locally jointly identifiable within this domain.

These results illustrate how FI can be used not only to quantify the precision with which parameters of stochastic systems may be inferred, but also to characterize how different physical mechanisms leave distinguishable statistical signatures in nonequilibrium fluctuations. In the present model, dynamical coupling and asymmetric athermal forcing affect the information geometry in qualitatively different ways. Extensions of this approach to experimentally sampled trajectories, rather than instantaneous probability distributions, and to more general non-Gaussian or nonlinear stochastic dynamics may provide further insight into parameter inference in nonequilibrium systems.

\section*{Acknowledgments}

WAMM acknowledges CNPq Grants No. 308560/2022-1 and 304863/2026-2.

\section*{Declaration of generative AI and AI-assisted technologies in the manuscript preparation process}
During the preparation of this work, the authors used ChatGPT 5.6 Sol for language editing, organization of the presentation, and critical review of the manuscript. The authors reviewed and edited the output as needed and take full responsibility for the content of the published article.

\appendix
%
%

%
%


\begin{thebibliography}{32}%
\makeatletter
\providecommand \@ifxundefined [1]{%
 \@ifx{#1\undefined}
}%
\providecommand \@ifnum [1]{%
 \ifnum #1\expandafter \@firstoftwo
 \else \expandafter \@secondoftwo
 \fi
}%
\providecommand \@ifx [1]{%
 \ifx #1\expandafter \@firstoftwo
 \else \expandafter \@secondoftwo
 \fi
}%
\providecommand \natexlab [1]{#1}%
\providecommand \enquote  [1]{``#1''}%
\providecommand \bibnamefont  [1]{#1}%
\providecommand \bibfnamefont [1]{#1}%
\providecommand \citenamefont [1]{#1}%
\providecommand \href@noop [0]{\@secondoftwo}%
\providecommand \href [0]{\begingroup \@sanitize@url \@href}%
\providecommand \@href[1]{\@@startlink{#1}\@@href}%
\providecommand \@@href[1]{\endgroup#1\@@endlink}%
\providecommand \@sanitize@url [0]{\catcode `\\12\catcode `\$12\catcode
  `\&12\catcode `\#12\catcode `\^12\catcode `\_12\catcode `\%12\relax}%
\providecommand \@@startlink[1]{}%
\providecommand \@@endlink[0]{}%
\providecommand \url  [0]{\begingroup\@sanitize@url \@url }%
\providecommand \@url [1]{\endgroup\@href {#1}{\urlprefix }}%
\providecommand \urlprefix  [0]{URL }%
\providecommand \Eprint [0]{\href }%
\providecommand \doibase [0]{http://dx.doi.org/}%
\providecommand \selectlanguage [0]{\@gobble}%
\providecommand \bibinfo  [0]{\@secondoftwo}%
\providecommand \bibfield  [0]{\@secondoftwo}%
\providecommand \translation [1]{[#1]}%
\providecommand \BibitemOpen [0]{}%
\providecommand \bibitemStop [0]{}%
\providecommand \bibitemNoStop [0]{.\EOS\space}%
\providecommand \EOS [0]{\spacefactor3000\relax}%
\providecommand \BibitemShut  [1]{\csname bibitem#1\endcsname}%
\let\auto@bib@innerbib\@empty
\bibitem [{\citenamefont {Shannon}(1948)}]{Shannon1948}%
  \BibitemOpen
  \bibfield  {author} {\bibinfo {author} {\bibfnamefont {C.~E.}\ \bibnamefont
  {Shannon}},\ }\href {\doibase
  https://doi.org/10.1002/j.1538-7305.1948.tb01338.x} {\bibfield  {journal}
  {\bibinfo  {journal} {Bell System Technical Journal}\ }\textbf {\bibinfo
  {volume} {27}},\ \bibinfo {pages} {379} (\bibinfo {year} {1948})}\BibitemShut
  {NoStop}%
\bibitem [{\citenamefont {Parrondo}\ \emph {et~al.}(2015)\citenamefont
  {Parrondo}, \citenamefont {Horowitz},\ and\ \citenamefont
  {Sagawa}}]{Parrondo2015}%
  \BibitemOpen
  \bibfield  {author} {\bibinfo {author} {\bibfnamefont {J.~M.~R.}\
  \bibnamefont {Parrondo}}, \bibinfo {author} {\bibfnamefont {J.~M.}\
  \bibnamefont {Horowitz}}, \ and\ \bibinfo {author} {\bibfnamefont
  {T.}~\bibnamefont {Sagawa}},\ }\href {\doibase 10.1038/nphys3230} {\bibfield
  {journal} {\bibinfo  {journal} {Nature Physics}\ }\textbf {\bibinfo {volume}
  {11}},\ \bibinfo {pages} {131} (\bibinfo {year} {2015})}\BibitemShut
  {NoStop}%
\bibitem [{\citenamefont {Granger}\ and\ \citenamefont
  {Kantz}(2011)}]{Granger2011}%
  \BibitemOpen
  \bibfield  {author} {\bibinfo {author} {\bibfnamefont {L.}~\bibnamefont
  {Granger}}\ and\ \bibinfo {author} {\bibfnamefont {H.}~\bibnamefont
  {Kantz}},\ }\href {\doibase 10.1103/PhysRevE.84.061110} {\bibfield  {journal}
  {\bibinfo  {journal} {Physical Review E}\ }\textbf {\bibinfo {volume} {84}},\
  \bibinfo {pages} {061110} (\bibinfo {year} {2011})}\BibitemShut {NoStop}%
\bibitem [{\citenamefont {Micadei}\ \emph {et~al.}(2013)\citenamefont
  {Micadei}, \citenamefont {Serra},\ and\ \citenamefont
  {C{\'e}leri}}]{Micadei2013}%
  \BibitemOpen
  \bibfield  {author} {\bibinfo {author} {\bibfnamefont {K.}~\bibnamefont
  {Micadei}}, \bibinfo {author} {\bibfnamefont {R.~M.}\ \bibnamefont {Serra}},
  \ and\ \bibinfo {author} {\bibfnamefont {L.~C.}\ \bibnamefont {C{\'e}leri}},\
  }\href {\doibase 10.1103/PhysRevE.88.062123} {\bibfield  {journal} {\bibinfo
  {journal} {Physical Review E}\ }\textbf {\bibinfo {volume} {88}},\ \bibinfo
  {pages} {062123} (\bibinfo {year} {2013})}\BibitemShut {NoStop}%
\bibitem [{\citenamefont {Linpeng}\ \emph {et~al.}(2022)\citenamefont
  {Linpeng}, \citenamefont {Bresque}, \citenamefont {Maffei}, \citenamefont
  {Jordan}, \citenamefont {Auffèves},\ and\ \citenamefont
  {Murch}}]{Linpeng2022}%
  \BibitemOpen
  \bibfield  {author} {\bibinfo {author} {\bibfnamefont {X.}~\bibnamefont
  {Linpeng}}, \bibinfo {author} {\bibfnamefont {L.}~\bibnamefont {Bresque}},
  \bibinfo {author} {\bibfnamefont {M.}~\bibnamefont {Maffei}}, \bibinfo
  {author} {\bibfnamefont {A.~N.}\ \bibnamefont {Jordan}}, \bibinfo {author}
  {\bibfnamefont {A.}~\bibnamefont {Auffèves}}, \ and\ \bibinfo {author}
  {\bibfnamefont {K.~W.}\ \bibnamefont {Murch}},\ }\href {\doibase
  10.1103/PhysRevLett.128.220506} {\bibfield  {journal} {\bibinfo  {journal}
  {Physical Review Letters}\ }\textbf {\bibinfo {volume} {128}} (\bibinfo
  {year} {2022}),\ 10.1103/PhysRevLett.128.220506}\BibitemShut {NoStop}%
\bibitem [{\citenamefont {Jacobs}(2012)}]{Jacobs2012}%
  \BibitemOpen
  \bibfield  {author} {\bibinfo {author} {\bibfnamefont {K.}~\bibnamefont
  {Jacobs}},\ }\href {\doibase 10.1103/PhysRevE.86.040106} {\bibfield
  {journal} {\bibinfo  {journal} {Physical Review E - Statistical, Nonlinear,
  and Soft Matter Physics}\ }\textbf {\bibinfo {volume} {86}} (\bibinfo {year}
  {2012}),\ 10.1103/PhysRevE.86.040106}\BibitemShut {NoStop}%
\bibitem [{\citenamefont {Sagawa}\ and\ \citenamefont
  {Ueda}(2010)}]{Sagawa2010}%
  \BibitemOpen
  \bibfield  {author} {\bibinfo {author} {\bibfnamefont {T.}~\bibnamefont
  {Sagawa}}\ and\ \bibinfo {author} {\bibfnamefont {M.}~\bibnamefont {Ueda}},\
  }\href {\doibase 10.1103/PhysRevLett.104.090602} {\bibfield  {journal}
  {\bibinfo  {journal} {Physical Review Letters}\ }\textbf {\bibinfo {volume}
  {104}},\ \bibinfo {pages} {090602} (\bibinfo {year} {2010})}\BibitemShut
  {NoStop}%
\bibitem [{\citenamefont {Sagawa}(2013)}]{Sagawa2013}%
  \BibitemOpen
  \bibfield  {author} {\bibinfo {author} {\bibfnamefont {T.}~\bibnamefont
  {Sagawa}},\ }\href {\doibase 10.1007/978-4-431-54168-4} {\emph {\bibinfo
  {title} {Thermodynamics of Information Processing in Small Systems}}}\
  (\bibinfo  {publisher} {Springer},\ \bibinfo {address} {Tokyo},\ \bibinfo
  {year} {2013})\BibitemShut {NoStop}%
\bibitem [{\citenamefont {Mancino}\ \emph {et~al.}(2018)\citenamefont
  {Mancino}, \citenamefont {Sbroscia}, \citenamefont {Roccia}, \citenamefont
  {Gianani}, \citenamefont {Somma}, \citenamefont {Mataloni}, \citenamefont
  {Paternostro},\ and\ \citenamefont {Barbieri}}]{Mancino2018}%
  \BibitemOpen
  \bibfield  {author} {\bibinfo {author} {\bibfnamefont {L.}~\bibnamefont
  {Mancino}}, \bibinfo {author} {\bibfnamefont {M.}~\bibnamefont {Sbroscia}},
  \bibinfo {author} {\bibfnamefont {E.}~\bibnamefont {Roccia}}, \bibinfo
  {author} {\bibfnamefont {I.}~\bibnamefont {Gianani}}, \bibinfo {author}
  {\bibfnamefont {F.}~\bibnamefont {Somma}}, \bibinfo {author} {\bibfnamefont
  {P.}~\bibnamefont {Mataloni}}, \bibinfo {author} {\bibfnamefont
  {M.}~\bibnamefont {Paternostro}}, \ and\ \bibinfo {author} {\bibfnamefont
  {M.}~\bibnamefont {Barbieri}},\ }\href {\doibase 10.1038/s41534-018-0069-z}
  {\bibfield  {journal} {\bibinfo  {journal} {npj Quantum Information}\
  }\textbf {\bibinfo {volume} {4}} (\bibinfo {year} {2018}),\
  10.1038/s41534-018-0069-z}\BibitemShut {NoStop}%
\bibitem [{\citenamefont {Pietzonka}\ and\ \citenamefont
  {Coghi}(2024)}]{Pietzonka2024}%
  \BibitemOpen
  \bibfield  {author} {\bibinfo {author} {\bibfnamefont {P.}~\bibnamefont
  {Pietzonka}}\ and\ \bibinfo {author} {\bibfnamefont {F.}~\bibnamefont
  {Coghi}},\ }\href {\doibase 10.1103/PhysRevE.109.064128} {\bibfield
  {journal} {\bibinfo  {journal} {Physical Review E}\ }\textbf {\bibinfo
  {volume} {109}},\ \bibinfo {pages} {064128} (\bibinfo {year}
  {2024})}\BibitemShut {NoStop}%
\bibitem [{\citenamefont {Fisher}(1925)}]{Fisher1925}%
  \BibitemOpen
  \bibfield  {author} {\bibinfo {author} {\bibfnamefont {R.}~\bibnamefont
  {Fisher}},\ }\href {\doibase doi:10.1017/S0305004100009580} {\bibfield
  {journal} {\bibinfo  {journal} {Mathematical Proceedings of the Cambridge
  Philosophical Society.}\ }\textbf {\bibinfo {volume} {22}},\ \bibinfo {pages}
  {700} (\bibinfo {year} {1925})}\BibitemShut {NoStop}%
\bibitem [{\citenamefont {Rice}(2007)}]{JARice_livro}%
  \BibitemOpen
  \bibfield  {author} {\bibinfo {author} {\bibfnamefont {J.~A.}\ \bibnamefont
  {Rice}},\ }\href@noop {} {\emph {\bibinfo {title} {Mathematical Statistics
  and data analysis}}},\ \bibinfo {edition} {3rd}\ ed.\ (\bibinfo  {publisher}
  {Cengage},\ \bibinfo {address} {Andover},\ \bibinfo {year}
  {2007})\BibitemShut {NoStop}%
\bibitem [{\citenamefont {Kay}(1993)}]{Kay_livro}%
  \BibitemOpen
  \bibfield  {author} {\bibinfo {author} {\bibfnamefont {S.~M.}\ \bibnamefont
  {Kay}},\ }\href@noop {} {\emph {\bibinfo {title} {Fundamentals of Statistical
  Signal Processing: Estimation Theory}}},\ Vol.~\bibinfo {volume} {1}\
  (\bibinfo  {publisher} {Prentice Hall PTR},\ \bibinfo {address} {New
  Jersey},\ \bibinfo {year} {1993})\BibitemShut {NoStop}%
\bibitem [{\citenamefont {Feitosa}\ and\ \citenamefont
  {Menon}(2002)}]{Feitosa2002}%
  \BibitemOpen
  \bibfield  {author} {\bibinfo {author} {\bibfnamefont {K.}~\bibnamefont
  {Feitosa}}\ and\ \bibinfo {author} {\bibfnamefont {N.}~\bibnamefont
  {Menon}},\ }\href {\doibase 10.1103/PhysRevLett.88.198301} {\bibfield
  {journal} {\bibinfo  {journal} {Physical Review Letters}\ }\textbf {\bibinfo
  {volume} {88}},\ \bibinfo {pages} {198301} (\bibinfo {year} {2002})}\BibitemShut
  {NoStop}%
\bibitem [{\citenamefont {Melo}\ \emph
  {et~al.}(2025{\natexlab{a}})\citenamefont {Melo}, \citenamefont
  {Queir{\'o}s}, \citenamefont {Soares-Pinto},\ and\ \citenamefont
  {Morgado}}]{Melo2025Geometric}%
  \BibitemOpen
  \bibfield  {author} {\bibinfo {author} {\bibfnamefont {P.~B.}\ \bibnamefont
  {Melo}}, \bibinfo {author} {\bibfnamefont {S.~M.~D.}\ \bibnamefont
  {Queir{\'o}s}}, \bibinfo {author} {\bibfnamefont {D.~O.}\ \bibnamefont
  {Soares-Pinto}}, \ and\ \bibinfo {author} {\bibfnamefont {W.~A.~M.}\
  \bibnamefont {Morgado}},\ }\href {\doibase 10.1088/1751-8121/add229}
  {\bibfield  {journal} {\bibinfo  {journal} {Journal of Physics A:
  Mathematical and Theoretical}\ }\textbf {\bibinfo {volume} {58}},\ \bibinfo
  {pages} {195001} (\bibinfo {year} {2025}{\natexlab{a}})}\BibitemShut
  {NoStop}%
\bibitem [{\citenamefont {Melo}\ \emph
  {et~al.}(2025{\natexlab{b}})\citenamefont {Melo}, \citenamefont {Iemini},
  \citenamefont {Soares-Pinto}, \citenamefont {Queir{\'o}s},\ and\
  \citenamefont {Morgado}}]{Melo2025SFI}%
  \BibitemOpen
  \bibfield  {author} {\bibinfo {author} {\bibfnamefont {P.~B.}\ \bibnamefont
  {Melo}}, \bibinfo {author} {\bibfnamefont {F.}~\bibnamefont {Iemini}},
  \bibinfo {author} {\bibfnamefont {D.~O.}\ \bibnamefont {Soares-Pinto}},
  \bibinfo {author} {\bibfnamefont {S.~M.~D.}\ \bibnamefont {Queir{\'o}s}}, \
  and\ \bibinfo {author} {\bibfnamefont {W.~A.~M.}\ \bibnamefont {Morgado}},\
  }\href {\doibase 10.1103/yjxw-p2zt} {\bibfield  {journal} {\bibinfo
  {journal} {Physical Review E}\ }\textbf {\bibinfo {volume} {112}},\ \bibinfo
  {pages} {014126} (\bibinfo {year} {2025}{\natexlab{b}})}\BibitemShut
  {NoStop}%
\bibitem [{\citenamefont {Cram\'er}(1999)}]{Cramer_livro}%
  \BibitemOpen
  \bibfield  {author} {\bibinfo {author} {\bibfnamefont {H.}~\bibnamefont
  {Cram\'er}},\ }\href@noop {} {\emph {\bibinfo {title} {Mathematical Methods
  of Statistics}}}\ (\bibinfo  {publisher} {Princeton University Press},\
  \bibinfo {year} {1999})\BibitemShut {NoStop}%
\bibitem [{\citenamefont {Rao}(1945)}]{Rao1945}%
  \BibitemOpen
  \bibfield  {author} {\bibinfo {author} {\bibfnamefont {C.~R.}\ \bibnamefont
  {Rao}},\ }\href@noop {} {\bibfield  {journal} {\bibinfo  {journal} {Bulletin
  of the Calcutta Mathematical Society}\ }\textbf {\bibinfo {volume} {37}},\
  \bibinfo {pages} {81} (\bibinfo {year} {1945})}\BibitemShut {NoStop}%
\bibitem [{\citenamefont {Filliger}\ and\ \citenamefont
  {Reimann}(2007)}]{FilligerReimann2007}%
  \BibitemOpen
  \bibfield  {author} {\bibinfo {author} {\bibfnamefont {R.}~\bibnamefont
  {Filliger}}\ and\ \bibinfo {author} {\bibfnamefont {P.}~\bibnamefont
  {Reimann}},\ }\href {\doibase 10.1103/PhysRevLett.99.230602} {\bibfield
  {journal} {\bibinfo  {journal} {Phys. Rev. Lett.}\ }\textbf {\bibinfo
  {volume} {99}},\ \bibinfo {pages} {230602} (\bibinfo {year}
  {2007})}\BibitemShut {NoStop}%
\bibitem [{\citenamefont {Argun}\ \emph {et~al.}(2017)\citenamefont {Argun},
  \citenamefont {Soni}, \citenamefont {Dabelow}, \citenamefont {Bo},
  \citenamefont {Pesce}, \citenamefont {Eichhorn},\ and\ \citenamefont
  {Volpe}}]{ArgunVolpe2017}%
  \BibitemOpen
  \bibfield  {author} {\bibinfo {author} {\bibfnamefont {A.}~\bibnamefont
  {Argun}}, \bibinfo {author} {\bibfnamefont {J.}~\bibnamefont {Soni}},
  \bibinfo {author} {\bibfnamefont {L.}~\bibnamefont {Dabelow}}, \bibinfo
  {author} {\bibfnamefont {S.}~\bibnamefont {Bo}}, \bibinfo {author}
  {\bibfnamefont {G.}~\bibnamefont {Pesce}}, \bibinfo {author} {\bibfnamefont
  {R.}~\bibnamefont {Eichhorn}}, \ and\ \bibinfo {author} {\bibfnamefont
  {G.}~\bibnamefont {Volpe}},\ }\href {\doibase 10.1103/PhysRevE.96.052106}
  {\bibfield  {journal} {\bibinfo  {journal} {Phys. Rev. E}\ }\textbf {\bibinfo
  {volume} {96}},\ \bibinfo {pages} {052106} (\bibinfo {year}
  {2017})}\BibitemShut {NoStop}%
\bibitem [{\citenamefont {Ito}\ and\ \citenamefont
  {Dechant}(2020)}]{ItoDechant2020}%
  \BibitemOpen
  \bibfield  {author} {\bibinfo {author} {\bibfnamefont {S.}~\bibnamefont
  {Ito}}\ and\ \bibinfo {author} {\bibfnamefont {A.}~\bibnamefont {Dechant}},\
  }\href {\doibase 10.1103/PhysRevX.10.021056} {\bibfield  {journal} {\bibinfo
  {journal} {Phys. Rev. X}\ }\textbf {\bibinfo {volume} {10}},\ \bibinfo
  {pages} {021056} (\bibinfo {year} {2020})}\BibitemShut {NoStop}%
\bibitem [{\citenamefont {Nascimento}\ and\ \citenamefont
  {Morgado}(2021)}]{Nascimento2021}%
  \BibitemOpen
  \bibfield  {author} {\bibinfo {author} {\bibfnamefont {E.~S.}\ \bibnamefont
  {Nascimento}}\ and\ \bibinfo {author} {\bibfnamefont {W.~A.~M.}\ \bibnamefont
  {Morgado}},\ }\href {\doibase 10.1088/1742-5468/abd027} {\bibfield  {journal}
  {\bibinfo  {journal} {J. Stat. Mech.}\ }\textbf {\bibinfo {volume} {2021}},\
  \bibinfo {pages} {013301} (\bibinfo {year} {2021})}\BibitemShut {NoStop}%
\bibitem [{\citenamefont {Nascimento}\ and\ \citenamefont
  {Morgado}(2022)}]{Nascimento2022}%
  \BibitemOpen
  \bibfield  {author} {\bibinfo {author} {\bibfnamefont {E.~S.}\ \bibnamefont
  {Nascimento}}\ and\ \bibinfo {author} {\bibfnamefont {W.~A.~M.}\ \bibnamefont
  {Morgado}},\ }\href {\doibase 10.1088/1751-8121/ac8c07} {\bibfield  {journal}
  {\bibinfo  {journal} {Journal of Physics A: Mathematical and Theoretical}\
  }\textbf {\bibinfo {volume} {55}},\ \bibinfo {pages} {395003} (\bibinfo
  {year} {2022})}\BibitemShut {NoStop}%
\bibitem [{\citenamefont {van Kampen}(2007)}]{van_Kampen-book}%
  \BibitemOpen
  \bibfield  {author} {\bibinfo {author} {\bibfnamefont {N.~G.}\ \bibnamefont
  {van Kampen}},\ }\href@noop {} {\emph {\bibinfo {title} {Stochastic Processes
  in Physics and Chemistry}}},\ \bibinfo {edition} {3rd}\ ed.\ (\bibinfo
  {publisher} {North-Holland},\ \bibinfo {address} {Amsterdam},\ \bibinfo
  {year} {2007})\BibitemShut {NoStop}%
\bibitem [{\citenamefont {Gardiner}(2004)}]{Gardiner}%
  \BibitemOpen
  \bibfield  {author} {\bibinfo {author} {\bibfnamefont {C.~W.}\ \bibnamefont
  {Gardiner}},\ }\href@noop {} {\emph {\bibinfo {title} {Handbook of Stochastic
  Methods for Physics, Chemistry, and the Natural Sciences}}},\ \bibinfo
  {edition} {3rd}\ ed.\ (\bibinfo  {publisher} {Springer International
  Publishing},\ \bibinfo {year} {2004})\BibitemShut {NoStop}%
\bibitem [{\citenamefont {Mestres}\ \emph {et~al.}(2014)\citenamefont
  {Mestres}, \citenamefont {Martinez}, \citenamefont {Ortiz-Ambriz},
  \citenamefont {Rica},\ and\ \citenamefont {Roldan}}]{MestresRoldan2014}%
  \BibitemOpen
  \bibfield  {author} {\bibinfo {author} {\bibfnamefont {P.}~\bibnamefont
  {Mestres}}, \bibinfo {author} {\bibfnamefont {I.~A.}\ \bibnamefont
  {Martinez}}, \bibinfo {author} {\bibfnamefont {A.}~\bibnamefont
  {Ortiz-Ambriz}}, \bibinfo {author} {\bibfnamefont {R.~A.}\ \bibnamefont
  {Rica}}, \ and\ \bibinfo {author} {\bibfnamefont {E.}~\bibnamefont
  {Roldan}},\ }\href {\doibase 10.1103/PhysRevE.90.032116} {\bibfield
  {journal} {\bibinfo  {journal} {Phys. Rev. E}\ }\textbf {\bibinfo {volume}
  {90}},\ \bibinfo {pages} {032116} (\bibinfo {year} {2014})}\BibitemShut
  {NoStop}%
\bibitem [{\citenamefont {Dinis}\ \emph {et~al.}(2016)\citenamefont {Dinis},
  \citenamefont {Mart{\'{\i}}nez}, \citenamefont {Rold{\'{a}}n}, \citenamefont
  {Parrondo},\ and\ \citenamefont {Rica}}]{DinisRica2016}%
  \BibitemOpen
  \bibfield  {author} {\bibinfo {author} {\bibfnamefont {L.}~\bibnamefont
  {Dinis}}, \bibinfo {author} {\bibfnamefont {I.~A.}\ \bibnamefont
  {Mart{\'{\i}}nez}}, \bibinfo {author} {\bibfnamefont {{\'{E}}.}~\bibnamefont
  {Rold{\'{a}}n}}, \bibinfo {author} {\bibfnamefont {J.~M.~R.}\ \bibnamefont
  {Parrondo}}, \ and\ \bibinfo {author} {\bibfnamefont {R.~A.}\ \bibnamefont
  {Rica}},\ }\href {\doibase 10.1088/1742-5468/2016/05/054003} {\bibfield
  {journal} {\bibinfo  {journal} {Journal of Statistical Mechanics: Theory and
  Experiment}\ }\textbf {\bibinfo {volume} {2016}},\ \bibinfo {pages} {054003}
  (\bibinfo {year} {2016})}\BibitemShut {NoStop}%
\bibitem [{\citenamefont {Bae}\ \emph {et~al.}(2021)\citenamefont {Bae},
  \citenamefont {Lee}, \citenamefont {Kim},\ and\ \citenamefont
  {Jeong}}]{BaeJeong2021}%
  \BibitemOpen
  \bibfield  {author} {\bibinfo {author} {\bibfnamefont {Y.}~\bibnamefont
  {Bae}}, \bibinfo {author} {\bibfnamefont {S.}~\bibnamefont {Lee}}, \bibinfo
  {author} {\bibfnamefont {J.}~\bibnamefont {Kim}}, \ and\ \bibinfo {author}
  {\bibfnamefont {H.}~\bibnamefont {Jeong}},\ }\href {\doibase
  10.1103/PhysRevE.103.032148} {\bibfield  {journal} {\bibinfo  {journal}
  {Phys. Rev. E}\ }\textbf {\bibinfo {volume} {103}},\ \bibinfo {pages}
  {032148} (\bibinfo {year} {2021})}\BibitemShut {NoStop}%
\bibitem [{\citenamefont {Squarcini}\ \emph {et~al.}(2022)\citenamefont
  {Squarcini}, \citenamefont {Solon}, \citenamefont {Viot},\ and\ \citenamefont
  {Oshanin}}]{Squarcini_2022}%
  \BibitemOpen
  \bibfield  {author} {\bibinfo {author} {\bibfnamefont {A.}~\bibnamefont
  {Squarcini}}, \bibinfo {author} {\bibfnamefont {A.}~\bibnamefont {Solon}},
  \bibinfo {author} {\bibfnamefont {P.}~\bibnamefont {Viot}}, \ and\ \bibinfo
  {author} {\bibfnamefont {G.}~\bibnamefont {Oshanin}},\ }\href {\doibase
  10.1088/1751-8121/aca4aa} {\bibfield  {journal} {\bibinfo  {journal} {Journal
  of Physics A: Mathematical and Theoretical}\ }\textbf {\bibinfo {volume}
  {55}},\ \bibinfo {pages} {485001} (\bibinfo {year} {2022})}\BibitemShut
  {NoStop}%
\bibitem [{\citenamefont {Ciliberto}\ \emph {et~al.}(2013)\citenamefont
  {Ciliberto}, \citenamefont {Imparato}, \citenamefont {Naert},\ and\
  \citenamefont {Tanase}}]{CilibertoImparatoTanaseNaert2013}%
  \BibitemOpen
  \bibfield  {author} {\bibinfo {author} {\bibfnamefont {S.}~\bibnamefont
  {Ciliberto}}, \bibinfo {author} {\bibfnamefont {A.}~\bibnamefont {Imparato}},
  \bibinfo {author} {\bibfnamefont {A.}~\bibnamefont {Naert}}, \ and\ \bibinfo
  {author} {\bibfnamefont {M.}~\bibnamefont {Tanase}},\ }\href {\doibase
  10.1088/1742-5468/2013/12/p12014} {\bibfield  {journal} {\bibinfo  {journal}
  {Journal of Statistical Mechanics: Theory and Experiment}\ }\textbf {\bibinfo
  {volume} {2013}},\ \bibinfo {pages} {P12014} (\bibinfo {year}
  {2013})}\BibitemShut {NoStop}%
\bibitem [{\citenamefont {Miller}(1974)}]{Miller_livro}%
  \BibitemOpen
  \bibfield  {author} {\bibinfo {author} {\bibfnamefont {K.~S.}\ \bibnamefont
  {Miller}},\ }\href@noop {} {\emph {\bibinfo {title} {Complex Stochastic
  Processes: An Introduction to Theory and Application}}}\ (\bibinfo
  {publisher} {Addison-Wesley Publishing Company, Inc.},\ \bibinfo {address}
  {Massachusetts},\ \bibinfo {year} {1974})\BibitemShut {NoStop}%
\bibitem [{\citenamefont {Malag\`{o}}\ and\ \citenamefont
  {Pistone}(2015)}]{MalagoPistone2015}%
  \BibitemOpen
  \bibfield  {author} {\bibinfo {author} {\bibfnamefont {L.}~\bibnamefont
  {Malag\`{o}}}\ and\ \bibinfo {author} {\bibfnamefont {G.}~\bibnamefont
  {Pistone}},\ }in\ \href {\doibase 10.1145/2725494.2725510} {\emph {\bibinfo
  {booktitle} {Proceedings of the 2015 ACM Conference on Foundations of Genetic
  Algorithms XIII}}},\ \bibinfo {series and number} {FOGA '15}\ (\bibinfo
  {publisher} {Association for Computing Machinery},\ \bibinfo {address} {New
  York, NY, USA},\ \bibinfo {year} {2015})\ p.\ \bibinfo {pages}
  {150–162}\BibitemShut {NoStop}%
\end{thebibliography}
\end{document}